\newcommand{\beq}{\begin{equation}}
\newcommand{\eeq}{\end{equation}}
\newcommand{\dd}{{\rm d}}
\newcommand{\fig}[1]{Fig.\,\ref{#1}}
\newcommand{\eqn}[1]{Eq.\,(\ref{#1})}
\newcommand{\sect}[1]{Sect.\,\ref{#1}}
\newcommand{\tab}[1]{Table\,\ref{#1}}
\newcommand{\appx}[1]{Appendix\,\ref{#1}}

\documentclass{aa}  

\usepackage{xcolor}
\usepackage{graphicx}
\usepackage{txfonts}
\usepackage{dsfont}
\usepackage{mhchem}
\usepackage{environ}
\usepackage{array}
\usepackage{hyperref}
\hypersetup{
    colorlinks,
    citecolor=blue,
    filecolor=blue,
    linkcolor=blue,
    urlcolor=blue
}
\usepackage{datetime}

\usepackage{soul}

\usepackage{orcidlink}

\definecolor{orange}{rgb}{1,0.5,0}
\definecolor{sred}{rgb}{.5,0,0}

\newcommand{\carbox}{\textsc{Carbox}}
\newcommand{\jax}{\textsc{jax}}
\newcommand{\lax}{\textsc{lax}}
\newcommand{\equinox}{\textsc{equinox}}
\newcommand{\diffrax}{\textsc{diffrax}}
\newcommand{\quadex}{\textsc{quadex}}
\newcommand{\optax}{\textsc{optax}}

\begin{document}

   \title{End-to-end differentiable retrieval of molecular spectra using hydrodynamics, chemistry, and radiative transfer}
   %automatic interpretation of spectral cubes with a differentiable 3d model

   %\subtitle{I. Applying the pipeline to a Larson-like collapse model}

   \author{T.~Grassi
          \inst{1,2}\fnmsep\thanks{Corresponding author, \email{tgrassi@mpe.mpg.de}}\orcidlink{0000-0002-3019-1077}
          \and
          G.~Vermari\"en\inst{3,4}\orcidlink{0000-0002-4346-5858}
          \and
          J.~E.~Pineda\inst{1}\orcidlink{0000-0002-3972-1978}
          \and
          S.~Spezzano\inst{1}\orcidlink{0000-0002-6787-5245}          
          \and
          S.~Bovino\inst{5,6,7}
          \and
          P.~Caselli\inst{1,2}\orcidlink{0000-0003-1481-7911}
          }

   \institute{Max-Planck-Institut f\"ur Extraterrestrische Physik, Giessenbachstra{\ss}e 1, 85748 Garching, Germany 
         \and
             ORIGINS Excellence Cluster, Boltzmannstra{\ss}e 2, 85748 Garching, Germany
          \and
          Leiden Observatory, Leiden University, PO Box 9513, 2300 RA Leiden, The Netherlands
          \and
          SURF, Amsterdam, The Netherlands
          \and
          Department of Chemistry, Sapienza University of Rome, P.le Aldo Moro 5, 00185 Rome, Italy
          \and
            Departamento de Astronom\'ia, Facultad Ciencias F\'isicas y Matem\'aticas, Universidad de Concepci\'on, Av. Esteban Iturra s/n Barrio Universitario, Casilla 160, Concepci\'on, Chile
            \and
INAF, Osservatorio Astrofisico di Arcetri, Largo E. Fermi 5, 50125, Firenze, Italy
             }

   \date{Received -; accepted -}

% \abstract{}{}{}{}{} 
% 5 {} token are mandatory
 
  \abstract
   {}
   {We aim to reproduce observed molecular line emission using a pipeline that couples hydrodynamics, chemistry, and radiative transfer capable of simultaneously optimizing all relevant physical and chemical parameters.}
   {We developed an end-to-end differentiable \jax{} pipeline consisting of a custom hydrodynamical code, a modified version of the differentiable chemical code \carbox{}, and a custom radiative transfer code. We tested the framework using controlled synthetic data.}
   {We demonstrate that the framework can recover the parameters of hydrodynamical shock models directly from molecular line spectra and optimize selected chemical reaction rate coefficients through gradient-based optimization. The differentiable formulation enables efficient optimization of the coupled physical and chemical model while preserving the full time-dependent evolution.}
   {}


   \maketitle
%
%-------------------------------------------------------------------
%--------------------------------------------------------------------
%--------------------------------------------------------------------
\section{Introduction}\label{sect:introduction}
Inferring the physical and chemical conditions of astronomical sources from molecular-line observations is a central challenge in astrophysics. Several methods have been proposed through the years, leveraging chemo-dynamical 1D models to determine source kinematic (e.g., \citealt{Keto2015,Sipila2018,Sipila2022} in prestellar cores), or chemical postprocessing techniques based on more complicated 3D (magneto)hydrodynamical MHD simulations (e.g., \citealt{FerradaChamorro2021,Jensen2023,Priestley2023,NavarroAlmaida2024, Narayan2025}).

More recently, the use of machine learning and statistical techniques has increased, including deep reinforcement learning \citep{Qiu2025}, interpretable techniques \citep{Heyl2023,AsensioRamos2024,Diop2024,Grassi2025,Vermarien2025a,Vermarien2025c}, standard deep neural networks methods \citep{Behrens2024,Kessler2025,Megias2025,Morisset2025}, Bayesian methods \citep{Makrymallis2014,Accurso2017,Maffucci2018,DeCeuster2023,Heyl2023b,Lin2025,Palaud2025}, linear algebra-based techniques \citep{Juvela1996,GomezGonzlez2016,deMijolla2024}, as well as autoencoders \citep{Portillo2020,Grassi2022,Sulzer2023,ShafaatMahmud2025}.

In addition to these methods, differentiable frameworks have recently emerged in the astrophysical domain, showing clear advantages of employing these techniques. For example, \textsc{ExoJAX} \citep{Kawahara2025,Kawahara2025b} is the first differentiable spectrum model for exoplanets and brown dwarfs that can compute cross-sections as functions of temperature and pressure, thereby minimizing interpolation errors in high-dispersion spectra. Additionally, differentiable hydrodynamic solvers for astrophysical environments have recently emerged, enabling efficient reverse-mode differentiation of time-dependent fluid evolution \citep{Kochkov2021,Storcks2024,Wang2024,Horowitz2025a,Horowitz2025b,Horowitz2026,Potluri2026,Storcks2026}, as well as for reaction-diffusion problems \citep{Pavlov2026}.

Concurrently, the development of differentiable astrochemical models, exemplified by \carbox{} \citep{Vermarien2025c}, has enabled large non-equilibrium chemical networks to be optimized using gradient-based methods. Likewise, \texttt{Ray-trax} \citep{Branca2025} demonstrated fully differentiable time-dependent radiative transfer with support for inverse problems and adjoint calculations.

Differentiable models have also been employed for problem inversion in circumstellar disks \citep{Kueny2026} and protoplanetary disks \citep{Yoshida2026} to retrieve spatial dust features, as well as dust surface density, temperature, and grain-size distribution.

Despite these advances in the machine-learning and differentiable simulation landscape, existing approaches usually focus on individual components of the physical problem, and yet no existing system offers a fully differentiable pipeline that consistently connects dynamical evolution, time-dependent chemistry, radiative transfer, and emergent spectra within a unified framework.
Such a capability is particularly attractive for contemporary astrophysical inference problems. Moreover, differentiability of the solution allows for the propagation of derivatives of the observable with respect to any parameter, avoiding surrogate models and computationally expensive finite-difference approximations. These features make it possible to solve inverse problems in which physical conditions are inferred directly from spectra.

In this work, we present the first fully differentiable astrophysical pipeline that combines one-dimensional hydrodynamics, time-dependent chemistry, radiative transfer, and spectral synthesis within a single end-to-end framework, although their applicability depends on the models' underlying physics. Implemented entirely in \jax{}, our pipeline supports automatic differentiation in forward and reverse modes and preserves gradient information throughout the calculation. This allows us to optimize physical parameters against spectra and facilitate local sensitivity analysis.

In \sect{sect:pipeline}, we will present the hydrodynamical code, our modified version of \carbox{}, and the custom radiative transfer code, and we will first test the \carbox{} differentiability to infer the chemical rate coefficients. In \sect{sect:optimization}, we will employ the complete pipeline to retrieve the ground truth parameters of a shock with chemistry and radiative transfer from the corresponding synthetic spectrum. In \sect{sect:limitations}, we will describe the limitations of our model, while in \sect{sect:conclusions}, we present the conclusions.

%--------------------------------------------------------------------
%--------------------------------------------------------------------
\section{Differentiable pipeline}\label{sect:pipeline}
The newly developed \jax{} \citep{Jax2018} differentiable pipeline consists of a chain of three modules: (i) a custom 1D hydrodynamic code, (ii) a time-dependent chemical solver consisting of a modified version of \carbox{} \citep{Vermarien2025c}, and (iii) a custom radiative transfer code. This pipeline is end-to-end differentiable, meaning that it is possible to obtain the partial derivatives of all the desired quantities, in particular, the derivatives of the output spectra produced by the radiative transfer, with respect to the hydrodynamic parameters (e.g., the initial conditions), the chemical parameters (e.g., a set of reaction rates), or the radiative transfer parameters (e.g., the collisional rates). This allows the optimization of every desired parameter with respect to the expected spectral output, as well as the local sensitivity analysis.

The hydrodynamical code can be interpreted as an operator $y_{\rm h}=\mathcal{H}(\theta_{\rm h})$, that given a set of parameters $\theta_{\rm h}$ (e.g., the initial conditions or the integration time) produces an output $y_{\rm h}$ (e.g., the gas distribution at a given time, or the velocity distribution). Analogously, the time-dependent chemical code is $y_{\rm c}=\mathcal{C}(y_{\rm h}, \theta_{\rm c})$, where the parameters $\theta_{\rm c}$ could be the reaction rate coefficients or the integration time, and the output $y_{\rm c}$ are the abundances of the desired chemical species at the given time. Finally, the radiative transfer operator is $I \equiv y_{\rm r}=\mathcal{R}(y_{\rm h}, y_{\rm c}, \theta_{\rm r})$, where the parameters $\theta_{\rm r}$ are the molecular data or each cell's microturbulence, while the output $I$ is the spectrum of one or more chemical species and one or more atomic/molecular transitions.

\begin{figure*}
\centering
    \includegraphics[width=0.9\textwidth]{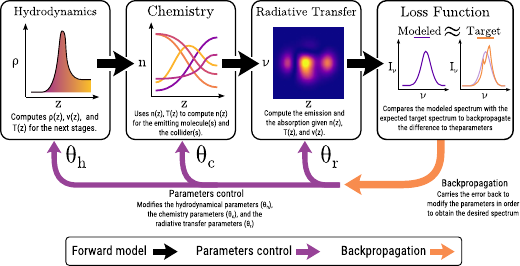}
    \caption{Sketch of the pipeline. The first step is to run a hydrodynamical code that evolves the density $\rho(z)$, velocity $v(z)$, and temperature $T(z)$. Then the chemical evolution produces the chemical species abundances $n(z)$. Finally, the radiative transfer produces the emission lines $I(\nu)$, or equivalently $I(\varv)$. Our code allows us to backpropagate the loss gradients with respect to the target spectra.}
        \label{fig:scheme}
\end{figure*}

Since all three operators are developed using \jax{}, thanks to automatic differentiation, through the inherent derivatives tree, we have simultaneously access to $\partial I/\partial \theta_{h}$, $\partial I/\partial \theta_{c}$, and $\partial I/\partial \theta_{r}$, i.e., we can differentiate the output spectra $I$ with respect to any parameter $\theta$ in the pipeline.

In addition to the forward pipeline, \jax{} allows for computing the backward pass, which is essential in applications such as constrained optimization, parameter inference, and sensitivity analysis, where gradients of a loss function with respect to initial conditions or model parameters are required. These gradients propagate information backward through the system's entire time evolution.

We will describe the details of each pipeline module in the next sections.

%--------------------------------------------------------------------
\subsection{Differentiable hydrodynamics}\label{sect:hydro}
We have developed a 1D hydrodynamical code which integrates the conservative form of the hydrodynamical equations for density ($\rho$), momentum ($\rho\varv$), and energy density ($E$), using a standard Riemann HLL solver \citep{Harten1983}. See \appx{appx:HLL} for further details.

For the aims of the present paper, we limit the code to have a single velocity component and no magnetic fields. These missing features are relevant, for example, in modeling oblique shocks, and the technical details of the implementations have already been discussed in \citet{Grassi2019}. Analogously, we do not resolve chemical evolution alongside the code, which will influence the temperature evolution via non-adiabatic cooling and heating and, in some cases, by affecting the fluid adiabatic index, mean molecular weight, and the non-ideal MHD coefficients. This presents an additional numerical challenge, more in terms of computational cost rather than development complexity (e.g., see \citealt{Grassi2014} and the discussion in \sect{sect:carbox}). Despite these shortcomings, the conclusions of this work remain unaffected. However, in the future, we will increase the complexity of this pipeline by including these features.

Although implementing the HLL Riemann solver presents no particular difficulties for automatic differentiation, the second-order Runge-Kutta method used to integrate the equation in time warrants further discussion due to the variable timestep. 

In fact, the timestep is determined by the Courant--Friedrichs--Lewy (CFL) condition \citep{Lewy1928}.
The maximum characteristic speed, considering all the cells \citep{Toro1994}, is
\begin{equation}
 \varv_{\max} = \max_i\left(|\varv_i|+c_{{\rm s},i}\right)\,,
\end{equation}
where \(c_{{\rm s},i}\) is the sound speed of the $i$th cell. The CFL timestep is then
\begin{equation}
 \Delta t_{\rm CFL} = C_{\rm CFL}\frac{\Delta z}{\varv_{\max}}\,,
\end{equation}
with $C_{\rm CFL}=0.1$ the Courant number and $\Delta z$ the cell size. The timestep is limited by a prescribed maximum value \(\Delta t_{\max}\), and by the remaining integration time, yielding
\begin{equation}
\Delta t = \min\left( C_{\rm CFL}\frac{\Delta z}{\varv_{\max}}, \Delta t_{\max}, t_{\rm end,h}-t \right)\,,
\end{equation}
where $t_{\rm end, h}$ is the desired total integration time of the hydrodynamics, so that the last term avoids overshooting the desired end of the simulation, and $\Delta t_{\rm max}=10^{-2}\,\Delta z$ is to avoid extremely large timesteps.

The whole procedure, as in the present form, produces an adaptive variable time step $\Delta t$.
Therefore, this implementation of the time-stepping is based on a ``while loop'' that exits when the condition $t=t_{\rm end,h}$ is satisfied. Since $\Delta t$ is not known in advance, the number of iterations is unpredictable.
 This is not a problem for standard forward simulations, and in \jax{} the forward pass can be implemented efficiently using \lax{} ``while loop''. However, reverse-mode automatic differentiation through such dynamically sized loops is not directly supported, since the backward pass requires access to the sequence of intermediate states.
For this reason, gradient evaluation must rely on an adjoint strategy based on checkpointing and recomputation of the forward trajectory. This approach avoids storing the full time history while enabling reverse-mode differentiation through simulations with adaptive, state-dependent time stepping. We used the \equinox{} implementation\footnote{\texttt{while\_loop} function from \texttt{equinox.internal} in \url{https://github.com/patrick-kidger/equinox}, commit \texttt{38e2086}.}, which elegantly solves the problem by storing only a set of intermediate checkpoints.

The code has been tested with a standard Sod's shock tube test using 128 cells \citep{Sod1978}, yielding the expected diffusivity produced by the HLL solver \citep{Grassi2019,Teyssier2002,Teyssier2006}.

Written in the present form, the code allows us to compute the partial derivatives with respect to any desired parameter. An example related to the Sod's shock tube test is to obtain the derivatives of the final density distribution with respect to the final integration time $t_{\rm end,h}$. In practice, we solve the forward problem up to a desired time, for example $t_{\rm end,h}=0.2$\,s, and then via the backward pass we obtain $\partial \rho(z)/\partial t_{\rm end,h}$, i.e., we obtain the derivative of the density with respect to the total integration time at every spatial position when $t=t_{\rm end,h}$. This can be extended to every output quantity that can be derived with respect to every parameter. We will leverage this property to enable effective parameter optimization.

%--------------------------------------------------------------------
\subsection{Differentiable chemistry}\label{sect:carbox}
For this stage of the pipeline, we need a differentiable time-dependent integrator. To this aim, we use the \carbox{} code \citep{Vermarien2025c}, developed in \jax{} and designed to be differentiable. The code is publicly available\footnote{\url{https://github.com/GijsVermarien/carbox/}, modified from commit \texttt{839b465}.}, and described in detail in the companion paper. This code solves the classic problem associated with a set of chemical reactions, i.e., solving the system of coupled ordinary differential equations
\beq\label{eqn:chemode}
    \frac{\dd n_i(t)}{\dd t} = P(n, k) - n_i L(n, k)\,,
\eeq
where $n_i$ is the number density of a given chemical species, $P(n, k)$ is the species production, as a function of all the abundances $n$ and all the reaction rate coefficients $k$, while $L$ is the destruction of the given species. This set of equations is integrated in time using a solver that supports stiffness, adaptive time-stepping, and, preferably, Jacobian sparsity \citep{Hindmarsh2005}. \carbox{} integrates the system of equations by using a differentiable solver from the library \diffrax{}\footnote{\url{https://github.com/patrick-kidger/diffrax}, commit \texttt{ae856ad}.} \citep{Kidger2022}, in particular, a Kvaerno integrator of order 3 or 5, depending on the stiffness \citep{Kvaerno2004}. Although this solver, when applied to large systems of differential chemical equations, is less efficient than well-established BDF solvers \citep{Hindmarsh2005}, the advantage is the capability of allowing backpropagation via a recursive checkpoints adjoint method, similar to the one described for the time ``while loop'' of the hydrodynamical solver.
In this way, the code is capable of differentiating the output with respect to the initial conditions and various parameters (e.g., temperature, cosmic-ray ionization rate, and visual extinction). The output is represented either by the final abundances\footnote{Note that $t_{\rm end}$ here is not the same as the hydrodynamical code.} $n(t=t_{\rm end})$, but also the complete evolution of the abundances in time, i.e., the trajectory $n(t)$ from $t=0$ to $t=t_{\rm end}$. 

For our pipeline, we modified \carbox{} to extend its differentiability with respect to individual rate equations. We assumed that the $i$th rate expression in the chemical network is controlled by three coefficients $\alpha_i$, $\beta_i$, and $\gamma_i$ and can have three main forms: (i) the canonic Arrhenius $k_i=\alpha_i (T / 300\,{\rm K})^\beta_i \exp(-\gamma_i /T)$ for generic gas-phase reactions, (ii) the cosmic-rays-induced reactions form $k_i=\alpha_i \zeta$, where $\zeta$ is the ionization rate of \ce{H2}, and (iii) $k_i=\alpha_i F_{\rm UV} \exp(-\gamma_i A_{\rm V})$ for the photochemical reactions, where $F_{\rm UV}$ is the radiation intensity in Draine's units, and $A_{\rm V}$ is the visual extinction. We limit the present set of rate equations to these gas-phase reactions, aware that other types of reactions exist (e.g., three-body and surface). We modified the code to retrieve derivatives with respect to any of the parameters mentioned here. For example, we can differentiate the abundance of a specific species at a given time with respect to a parameter of a given reaction rate, e.g., $\partial n_{\ce{HCO+}}(t=10\,{\rm kyr}) /\partial \alpha_{5}$, where the abundance of \ce{HCO+} at 10\,kyr is differentiated with respect to $\alpha_{5}$, one of the parameters of the 5th reaction in the chemical network. This can be calculated for any combination of outputs and parameters.

These modifications enable optimization of reaction rates with respect to a specific target. We tested the capability of our code of leveraging automatic differentation by predicting the rate coefficient of a chemical network using a given set of time-dependent abundances $n(t; T, \zeta)$ at different temperatures and cosmic ray ionization rates. Instead of optimizing a single set of $n(t)$ trajectories (i.e., at a single $T$ and $\zeta$), we run multiple instances of \carbox{} on a grid of $T$ and $\zeta$ values all at the same time. This is possible thanks to vectorization via \texttt{jax.vmap}, and we optimize all of $\alpha$, $\beta$, and $\gamma$ together. This is equivalent to optimizing the rate coefficients simultaneously across multiple temperatures and cosmic-ray ionization values.

The network employed is relatively small, consisting of 25 reactions (4 cosmic-ray-induced, 4 photochemical, and the rest standard gas-phase), and is based on \citet{Bialy2015}, including their water formation path starting with cosmic-ray-induced oxygen ionization and the CO/\ce{HCO+} formation route (see \appx{appx:network}). This is designed to capture interpretable chemical effects while remaining relatively time- and memory-efficient. Although the network is relatively simple, it already has 63 free parameters to be optimized simultaneously. We are aware that this network is not state-of-the-art and is used here only for illustrative purposes.

We note that the optimization can reproduce most of the target values for $\alpha$, $\beta$, and $\gamma$, except for some values that do not contribute to the modeled abundances over the selected range of $T$ and $\zeta$ and are therefore unconstrained. Details on the optimization method can be found in \appx{appx:testcarbox}.

The code's ability to perform simultaneous vectorized optimization across models with different parameters is crucial for the application in \sect{sect:optimization}.
 
%--------------------------------------------------------------------
%--------------------------------------------------------------------
\subsection{Differentiable radiative transfer}\label{sect:rt}
In order to predict the emission from the simulated region, we developed a custom radiative transfer code written in \jax{}, similar to the one presented in \cite{Levis2025}, but including level-population calculations. We use the spectral data and collisional rate coefficients from the \textsc{Lamda} database \citep{Schoier2005}, in particular the energy of the $i$th excited levels $E_i$, their radiative coefficients $A_{ij}$ and collisional coefficients $C_{ij}(T)$ between the $i$th and $j$th levels, the latter for a set of available collisional partners. Since the temperature is a function of the position $z$, the collisional rate coefficients are log-interpolated from the database using an interpolator vectorized in the $z$ direction. These parameters allow us to construct a set of differential equations similar to \eqn{eqn:chemode}, but for the excited levels, where the $i$th level is populated by $\sum_k C_{ji}\,n_j\,n_k$, where $k$ is the list of colliders, and by $A_{ji}\,n_j$ when $j>i$. Analogously, the level is depopulated by $-\sum_k C_{ij}\,n_i\,n_k$ and $-A_{ij}\,n_i$ when $i>j$. By assuming equilibrium ($\dd n_i/\dd t=0$) and imposing number density conservation ($\sum_in_i=n_{\rm tot}$), we solve the linear system to obtain the level populations $n_i$ (e.g., see \citealt{Maio2007}). This operation is vectorized to compute the level population over the full simulation domain simultaneously, i.e., $n_i(z)$, a matrix of $N_z\times N_{\rm L}$ elements, where $N_z$ is the number of cells, and $N_{\rm L}$ the number of excited levels.

The level population is used to compute the emission of the transition at $\nu_0$ from the $i$th to the $j$th level,
\beq
 j_\nu(z)= \frac{h \nu_0}{4\pi} A_{ij} n_i(z)\,\phi\left(\nu, z\right) \,,
\eeq
where $h$ is Planck's constant, and the line shape is 
\beq
 \phi(\nu, z) = \frac{c}{a(z) \nu_0 \sqrt{\pi}}\exp\left[-\frac{c^2(\nu-\hat\nu_0)^2}{\nu_0^2 a(z)^2}\right]\,,
\eeq
with $c$ the speed of light, $\hat \nu_0(z) = \nu_0 (1 - \varv(z) / c)$ takes into account the Doppler effect produced by the moving gas cell\footnote{In our case, we have a single velocity component $\varv\equiv\varv_z$, so there is no need to deproject the component with respect to the observer.}, $\nu_0$ is the rest frame frequency of the given transition, and
\beq
 a(z) = \sqrt{\varv^2_{\rm turb}(z) + \frac{2 k_{\rm B} T(z)}{\mu}}\,,
\eeq
where $\varv_{\rm turb}$ is the turbulence parameter, and $\mu$ is the chemical species mean molecular weight\footnote{Note: in units of mass (g), not amu.}.

Analogously, the absorption is 
\beq
 \alpha_{\nu}(z) = \frac{h\nu_0}{4\pi} \left(B_{lu} n_l - B_{ul} n_i\right)\,\phi(\nu, z)\,
\eeq
where $B_{ul}=A_{ul} c^2 / (2 h\nu_0^3)$ and $B_{lu}g_l=B_{ul}g_u$ are defined assuming upper ($u$) and lower ($l$) level transitions, where $g$ are the statistical weights of the levels.

The total absorption from the observer to a specific position $z$ is given by the cumulative integral\footnote{We use the function of the library \quadex{} that is \jax{}-friendly, supports automatic differentiations, and is easily vectorizable. See \url{https://github.com/f0uriest/quadax}, commit \texttt{2273ba8}.} along the spatial dimension
\beq
 \tau_\nu(z) = \int_0^z \alpha_\nu(z')\,\dd z'\,,
\eeq
and therefore the final spectrum is obtained by integrating the full domain of size $L$
\beq
 I(\nu) = \int_0^L j_{\nu}(z)\,e^{-\tau_\nu(z)} \dd z\,,
\eeq
where the output along the line of sight is in cgs units is erg\,s$^{-1}$\,cm$^{-2}$\,Hz$^{-1}$, i.e., units of $10^{23}$\,Jy. The code can include background radiation, not employed in this work. All the operations mentioned above are vectorized along the $z$ coordinate, but also along the frequency coordinate, so that we can compute $I$ for all the molecular transitions at the same time. The code internally uses frequency units, but the output spectra are functions of velocity, i.e., $\varv=c(\nu_0 - \nu) / \nu_0$.

The code is implemented to take as input the name of the molecule, the list of desired colliders, the spatial distribution of the molecule number density $n_{\rm mol}(z)$, the number density distribution of the colliders $n_k(z)$, the gas velocity $\varv(z)$, its temperature $T(v)$, and the turbulence $\varv_{\rm turb}(z)$. The result is a set of spectra $I(\varv)$, one for each chemical species transition, where $\varv=0$ corresponds to the rest frame, i.e., where $\nu=\nu_0$. The spectral resolution $\Delta \varv$ can also be chosen. The output spectrum is differentiable with respect to all the aforementioned parameters, except $\Delta\varv$.

The code has been benchmarked against the PDR benchmark results \citep{Rollig2007} for level populations and the emission of C, \ce{C+}, and O lines, using \textsc{Cloudy} as reference data \citep{Ferland2017}. The emission and absorption part has been tested against LOC \citep{Juvela1997,Juvela2020}. See \appx{appx:testrt} for further details.

\subsubsection{Optically thick case}
In this specific setup, we omit the contributions of the radiation-induced transitions to the level population. Their rate coefficients depend on the $B_{lu}$ and $B_{ul}$ coefficients but also on the radiation in the specific cell, or better, on the integral of the product of the line shape and the radiation field, i.e., $\int \phi(\nu, z) j_\nu( z)\,\dd\nu$. Although this is a local quantity, the radiation field in a cell $j_\nu(z)$ depends on the opacity and emission of the other cells, i.e., on the population of the excited levels in each cell and the opacity between cells. This suggests that we need to use an iterative procedure to achieve convergence of the level population. There are several techniques to solve this problem (e.g., see \citealt{vanZadelhoff2002,vanderTak2007,Dullemond2012}), but they all rely on an unknown number of iterations, which complicates the storage of the history of reverse-mode differentiation used to leverage differentiability, as we reported in \sect{sect:hydro}.

For testing, we implemented the full calculation using pure Lambda iteration without any specific accelerator, such as ALI or MALI \citep{Rybicki1991,Rybicki1992}. This is possible because the code reaches the convergence relatively quickly, in terms of actual GPU time, thanks to the extensive vectorization of the code to compute the opacity between the $i$th and the $j$th cell $\Delta\tau_{ij}(\nu)$, while the total radiation in each cell is computed with the \texttt{jax.einsum}, again leveraging complete vectorization. We solved the convergence issue of reverse differentiation using the checkpoints method, as in \sect{sect:hydro}.
We will employ the full radiative transfer code in future works.

%--------------------------------------------------------------------
%--------------------------------------------------------------------
\section{End-to-end optimization}\label{sect:optimization}
The complete pipeline allows us to differentiate the output spectrum $I(\varv)$ with respect to any parameter included in the hydrodynamical, chemical, and radiative transfer codes. Thanks to this capability, analogously to the optimization test with \carbox{}, we can optimize the spectrum $I(\varv)$ produced by the radiative transfer code to the desired parameters.

As an example, we model a 1D shock in a box of size $L=10^4$\,au with 128 grid elements and initial jump conditions (left and right) at $z=L/2$ as $(n_{\rm L}, n_{\rm R})$=$(10^3, 4\times10^3)$\,cm$^{-3}$, $(\varv_{\rm L}, \varv_{\rm R})$=$(1, 0)$\,km\,s$^{-1}$, $(T_{\rm L}, T_{\rm R})$=$(10^3, 10)$\,K, i.e., a hot moving gas impacting on a denser, cold, static gas. The gas is assumed ideal with adiabatic index $\gamma=1.4$ and mean molecular weight $\mu=2.33\,m_{\rm p}$, where $m_{\rm p}$ is the proton mass. The simulation is stopped at $t=10^3$\,yr. 

The final quantities $n(z)$ and $T(z)$ are provided to \carbox{}, that assuming $\zeta=10^{-17}$\,s$^{-1}$, $A_{\rm V}=2$\,mag, and initial conditions proportional to the total density as $n_{\ce{H2}}(z)=n(z)$, $n_{\rm C}(z)=10^{-4}n(z)$, and $n_{\rm O}(z)=3\times10^{-4}n(z)$ \citep{Rollig2007}, integrates the system of differential equations to $t_{\rm end}=10^5$\,yr. The chemical network is the same as that used in \sect{sect:carbox}, and the aim here is to mimic a chemical post-processing operation, since the chemistry is calculated after the complete hydrodynamic integration rather than during each hydrodynamical step via operator splitting.

The output quantity $n_{\ce{HCO+}}(z)$ is provided to the radiative transfer code, alongside $T(z)$ and $\varv(z)$ from the hydrodynamical code. The turbulence is set to $\varv_{\rm turb}(z)=0.07$\,km\,s$^{-1}$, constant in the whole domain. We employ this rather small value to have sharper spectral features in the output spectrum (cf. \citealt{Pineda2026}). The radiative transfer mimics an observation along the $z$ component of the shock, considering the transition $J=1\to 0$ of \ce{HCO+} ($J=0$ is the ground level), with $\nu_0=89.189$\,GHz \citep{Muller2005}. The spectrum spans $4$\,km\,s$^{-1}$ using 100 spectral channels. The chemical code also provides $n_{\ce{H2}}(z)$, which is employed to compute ortho-\ce{H2} and para-\ce{H2} colliders, considering an ortho-to-para ratio of 3, and we use the data from the \textsc{Lamda} database\footnote{\url{https://home.strw.leidenuniv.nl/~moldata/datafiles/hco+.dat}} \citep{Schoier2005}. We do not include any mock observational noise. Hence, the molecule can be observed regardless of the obtained abundance. However, the hydrodynamical and chemical evolutions play a significant role in the final observed abundance, thereby providing an effective benchmark for our pipeline.

In this optimization test, we want to obtain the exact shape of the \ce{HCO+} spectrum in the velocity space $I(\varv)$, by optimizing the parameters $n_{\rm L}$, $T_{\rm L}$, and $\varv_{\rm L}$ from the hydrodynamic module, $t_{\rm end}$, $\alpha_{18}$, and $\beta_{18}$ from the chemistry, and $v_{\rm turb}$ from the radiative transfer. In practice, we assume we do not know the left characteristics of the shock's initial conditions ($n_{\rm L}$, $T_{\rm L}$, and $\varv_{\rm L}$), the chemical post-processing integration time ($t_{\rm end}$) and the rate coefficients for $\ce{HCO+}\,+\,\ce{e-}\to\ce{CO}\,+\,\ce{H}$, namely $k(T)=\alpha_{18}(T/300\,{\rm K})^{\beta_{18}}$. Finally, we do not know the amount of microturbulence $v_{\rm turb}$ that determines the line shape. This set of parameters samples unknowns from each module of the pipeline.

We want to minimize the $L_2$-norm of the difference of the target and optimized $I(\varv)$, normalized to the maximum of the target, and we also normalize the input parameters in the logarithmic space. The Adam optimizer \citep{Kingma2014} is from the \jax{} optimization library \optax{} \citep{Optax2020} and has a learning rate of $0.03$, and the other parameters are set to their defaults. This learning rate is rather aggressive, but despite some initial oscillations, it recovers the results in a few iterations (see \appx{appx:loss}). We uniformly randomize in the logarithmic space the initial conditions in the ranges of $n_{\rm L}=[30$, 3$\times10^4$]\,cm$^{-3}$, $\varv_{\rm L}=[30, 3\times10^{5}$]\,cm\,s$^{-1}$, $T_{\rm L}=[30, 3\times10^{3}$]\,K,  $t_{\rm end}=[10^2, 10^6]$\,yr, $\varv_{\rm turb}$=[0.01 ,0.5]\,km\,s$^{-1}$, $\alpha_{18}=[10^{-8}, 10^{-4}]$\,cm$^3$\,s$^{-1}$, and $\beta_{18}$=[-2,  2], the last one normalized in linear space.

Not all the initial conditions produce effective convergence, suggesting that the curvature of the parameter hyperspace is not always positive. In theory, it could be possible to explore the eigenvalues of the corresponding Hessian matrix to determine the curvature, but the checkpointing on the backward differentiation used for the ``while loop'' in the hydrodynamics discussed in \sect{sect:hydro} is not designed to support second-order reverse mode.

\begin{figure*}
\centering
    \includegraphics[width=0.9\textwidth]{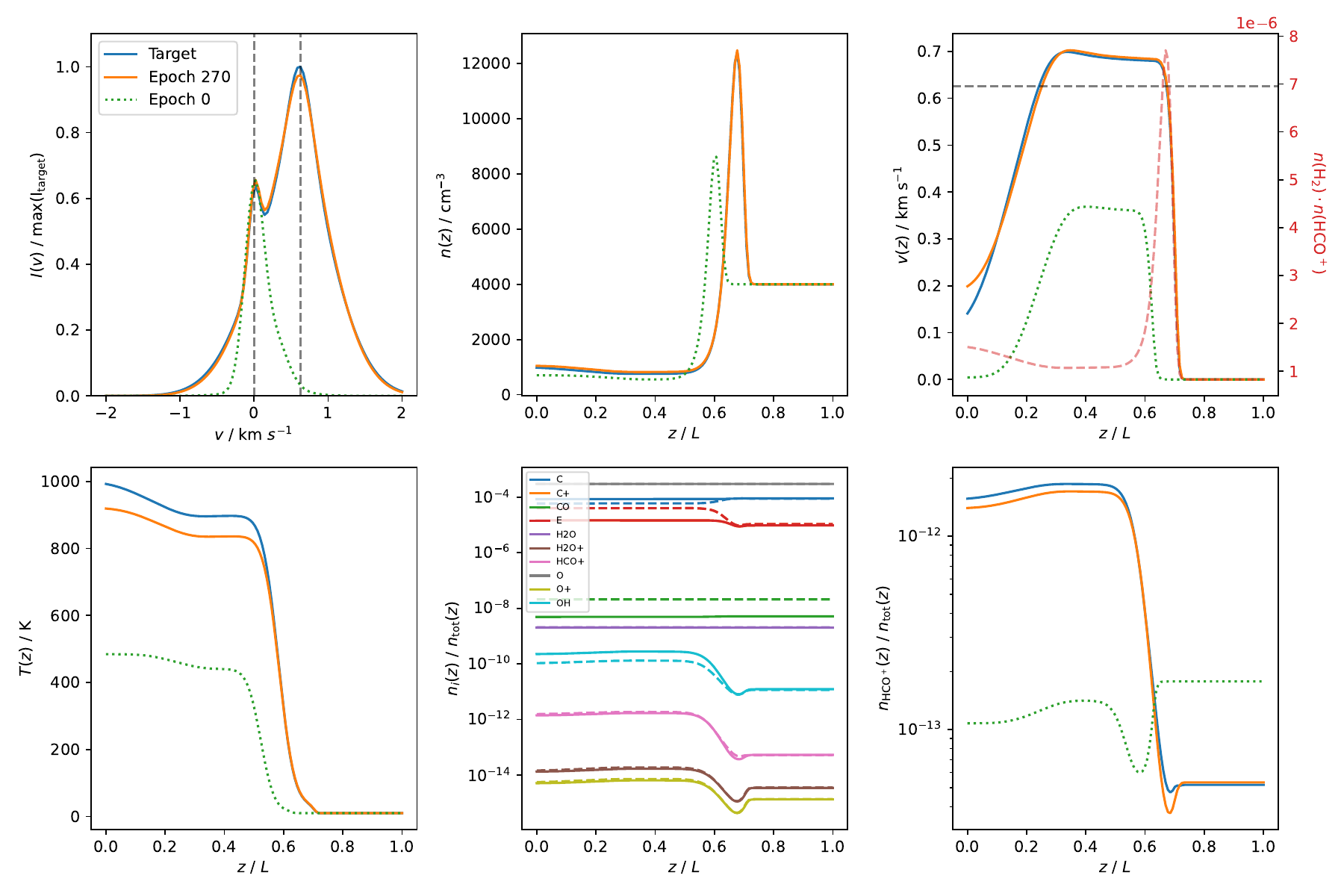}
    \caption{Comparison of the target quantities (solid blue), evaluated at epoch~0 (dotted green), and at epoch~270 (solid orange). We show the \ce{HCO+} ($1\to0$) emission normalized by the maximum of the target emission profile (top left), the density profile (top center), the velocity profile (top right, left black $y$-scale), the $n(\ce{HCO+})\cdot n(\ce{H2})$ product using the densities of the target (top right, right red $y$-scale), the temperature profile (bottom left), a subset of the chemical species relative abundances (center bottom), and the abundance of \ce{HCO+} (bottom right). The vertical dashed lines in the top-left panel show the peaks of the two components. In the top right panel, the horizontal dashed line corresponds to the velocity of the main feature at $\varv\approx0.6$\,km\,s$^{-1}$ in the top-left panel. The bottom center panel shows only the target (solid lines) and the epoch~270 quantities (dashed lines).  Note that the horizontal axis of the top left panel is velocity ($\varv$), while for the other panels is the spatial coordinate in units of the simulation box length ($z/L$).}
        \label{fig:optimization}
\end{figure*}

The solution reported in \fig{fig:optimization} is found after 270~epochs\footnote{Approximately 6 minutes on the NVIDIA Quadro RTX 4000, year 2018.}, but a reasonably accurate fit is found after around 60~iterations, and the remaining iterations are used to fine-tune the solution (see \appx{appx:loss}). The synthetic spectrum is well reproduced, including the profiles of the key quantities, proving that the pipeline is relatively robust and allows backward differentiation. The peak in the spectrum (first panel, top row) maps the main velocity components at approximately 0.6\,km\,s$^{-1}$ and the second peak on the left side is determined by the static gas from 0.75\,$z/L$ to $z/L$ (last panel, top row). The intensity of these two features maps the tracer (last panel, bottom row) and the colliders' abundances (top panel, central row, since \ce{H2} is the main constituent of the gas). In the top right panel, it should be noted that the 0.6\,km\,s$^{-1}$ feature (horizontal dashed line) is determined by the abundance of the tracer \ce{HCO+} and the collider \ce{H2}, as shown by the dashed red curve indicating the product $n(\ce{HCO+})\cdot n(\ce{H2})$.

For comparison, we report the solution at the first epoch in \fig{fig:optimization}, i.e., the solution with randomized parameters (dotted lines, except in the center panel, bottom row, omitted for clarity). We note that the initial abundance of \ce{HCO+} is far from the expected solution after one iteration, suggesting that the optimization of the chemistry might play a crucial role in the resulting spectrum. We also note that the optimized spectrum closely matches the target spectrum, whereas the other quantities show some discrepancies. This indicates that there are some degeneracies among the quantities, or that the spectrum is not very sensitive to them. An example is the abundances of CO and electrons (solid and dashed green and red lines in the lower central panel), which show a small discrepancy, indicating that their exact abundances are not relevant to the final spectrum. 

In \tab{tab:parameters}, the expected parameter values are all reproduced ($n_{\rm L}$, $T_{\rm L}$, $\varv_{\rm L}$, $\varv_{\rm turb}$, $\alpha_{18}$ and $\beta_{18}$), suggesting that they are all relevant to produce the target output. The only parameter that is relatively different is  $t_{\rm end}$, which is smaller than the expected target value. This is because, in this specific example, the abundance of \ce{HCO+} relevant to reproducing the spectra reaches the expected values after only $5\times10^3$~yr, and therefore the optimization does not need to evolve the chemistry for longer.

\begin{table}
    \caption{Target and optimized parameters after 270~epochs.}
    \centering
    \begin{tabular}{lccl}
        \hline
        Parameter& Target & Optimized & Units\\
        \hline
$n_{\rm L}$ & 1.0(3) & 9.5(2) & cm$^{-3}$\\
$v_{\rm L}$ & 1.0(4) & 1.2(4) & cm\,s$^{-1}$\\
$T_{\rm L}$ & 1.0(3) & 1.0(3) & K\\
$t_{\rm end}$ & 1.0(5) & 5.5(3) & yr\\
$\varv_{\rm turb}$ & 7.0(3) & 7.5(3) & cm\,s$^{-1}$\\
$\alpha_{18}$ & 2.8(-7) & 4.7(-7) & cm$^{3}$\,s$^{-1}$\\
$\beta_{18}$ & -0.69 & -0.59 & -\\
        \hline
    \end{tabular}
\tablefoot{$a(b)=a\times10^b$.}\label{tab:parameters}
\end{table}

% ---------------------------
\subsection{Local sensitivity analysis}

\begin{figure}
\centering
    \includegraphics[width=0.48\textwidth]{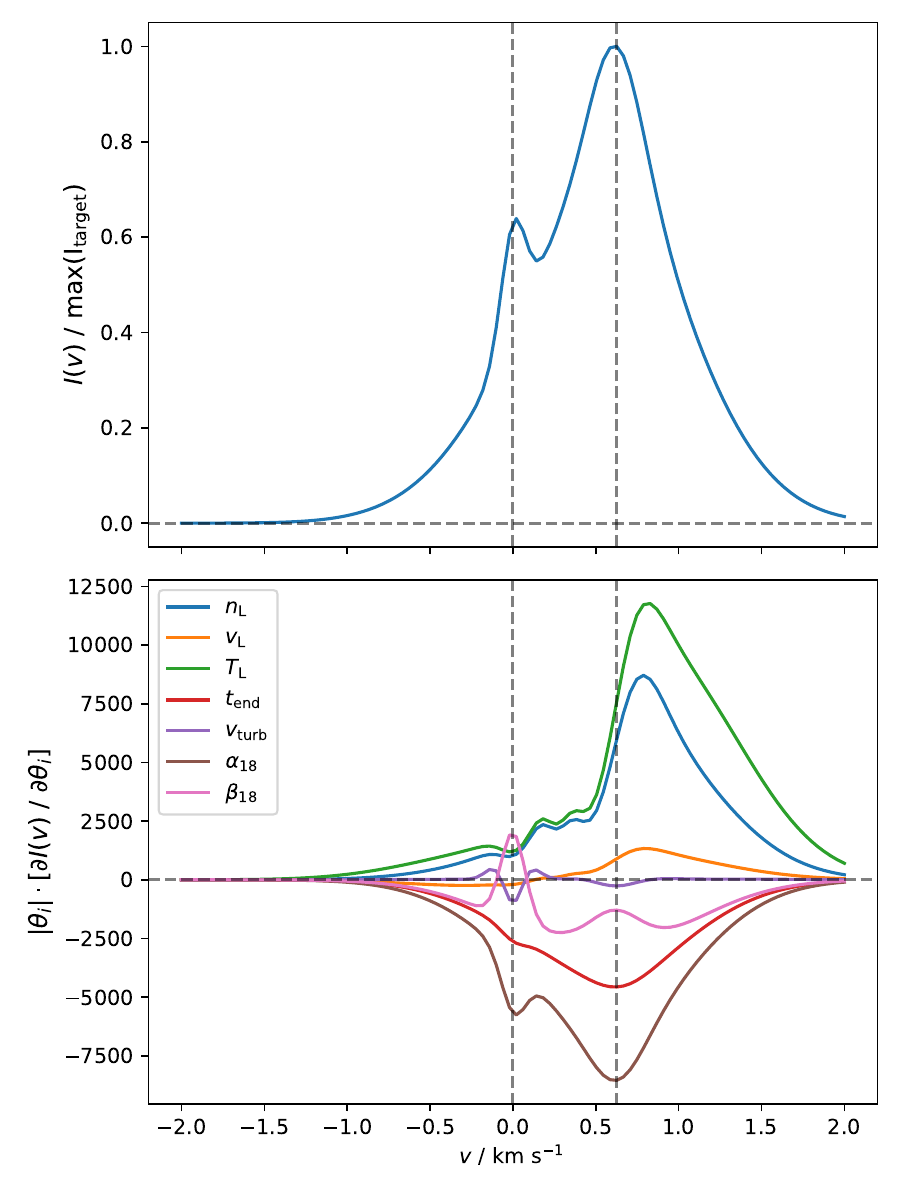}
    \caption{Top panel: the \ce{HCO+} ($1\to0$) emission normalized by the maximum of the target emission profile (same as top left panel of \fig{fig:optimization}). Bottom panel: the partial derivatives of the parameters multiplied by the value of the parameters evaluated as for the solution in the top panel, illustrating the sensitivity of the different
    parameters. In both panels, the dashed vertical lines indicate the position of the two peaks.}
        \label{fig:jacobian}
\end{figure}

The impact of the various parameters can be quantified by leveraging the intrinsic characteristics of automatic differentiation. Since, by construction, we have access to the partial derivatives, we can measure the local sensitivity with respect to the parameters. It is possible to compute the sensitivity to all the 100 spectral velocity channels with respect to every parameter with the Jacobian, e.g., $\partial I(\varv)/\partial n_{\rm L}$ or $\partial I(\varv)/\partial \alpha_{18}$. In our pipeline, due to the implementation of adjoint checkpoints in the hydrodynamic while loop, we cannot use the forward Jacobian-vector product (\texttt{jvp}), instead, we use the Jacobian evaluated row-by-row via reverse-mode automatic differentiation (\texttt{jacrev}). This will slightly reduce the computational efficiency, but it allows us to evaluate the derivatives reported in \fig{fig:jacobian}, where in the upper panel we show the optimized spectra for reference (same as in the first panel of \fig{fig:optimization}), and in the lower panel the corresponding derivatives. As a quantity for local sensitivity, we use $|\theta|\cdot\partial I(\varv)/\partial \theta$, which allows us to have a scaled partial derivative independent of the order of magnitude of the actual parameter value $\theta$.

We note that the feature in the larger peak around $0.6$\,km\,s$^{-1}$ positively correlates with the gas density and the temperature at the left interface ($n_{\rm L}$ and $T_{\rm L}$), suggesting that the intensity of the emitted line is controlled by the upstream density (more material, more emission) and the temperature (higher temperature, more populated excited levels). Analogously, the left interface velocity ($\varv_{\rm L}$) positively correlates with the peak intensity, indicating that increasing the upstream velocity will produce a brighter line peak (more material at the peak velocity).

It is worth noting that the position of the peak in the upper panel does not correspond to a peak in the derivatives in the lower panel, but they are shifted to higher velocity (see the vertical dashed line at  $0.6$\,km\,s$^{-1}$). This is because larger hydrodynamical parameters are expected to shift the emission peak to higher velocities (due to greater pressure or momentum). Conversely, the maxima/minima of the derivative of the non-hydrodynamical parameters correspond to the peaks because they do not affect the velocity of the fluid, but only the abundance of the tracer.

In fact, in \fig{fig:jacobian}, the chemical integration time $t_{\rm end}$ inversely correlates with the emission intensity, indicating that a longer chemical evolution will reduce the abundance of \ce{HCO+}. The parameter $\alpha_{18}$ controls the recombination of \ce{HCO+}, explaining the negative derivative over the entire domain. The slope of the rate coefficient temperature dependence is controlled by $\beta_{18}$ as $k(T)\propto T^{\beta_{18}}$. Higher $\beta_{18}$ means favoring the destruction of \ce{HCO+} at higher temperatures, explaining the negative derivative for the highest peak (it is produced in the hotter region $T\gtrsim800$\,K, see \fig{fig:optimization}), and the positive derivative for the zero-velocity peak (produced when $T\approx10$\,K).

Finally, the microturbulence velocity $\varv_{\rm turb}$ determines the shape of the lines, in particular, it plays a relevant role for the width of the line with zero velocity. This can be noted by the sign inversion of the purple line in \fig{fig:jacobian} around $\varv=0$\,km\,s$^{-1}$. It means that increasing $\varv_{\rm turb}$ increases the ``sides'' of the line, reducing the height of the central peak (cf. \citealt{Grassi2026}). This effect is slightly visible for the highest peak at $\varv\approx0.6$\,km\,s$^{-1}$.

%--------------------------------------------------------------------
%--------------------------------------------------------------------
\section{Limitations}\label{sect:limitations}
The principal limitation is the relatively low dimensionality of the problem, which reduces the actual impact on memory and realistic computational intensity. In particular, the hydrodynamical code is a 1D, single-velocity-component, no-magnetic-field, uniform Cartesian grid with a relatively diffusive solver (HLL). However, the main implementation challenge is backpropagating derivatives, particularly the uncertainty arising from the unknown number of integration steps due to adaptive time-stepping. The latter problem has been solved with checkpointing.

The chemical part using \carbox{} has a similar issue, i.e., we use a relatively small chemical network that captures some key chemical features, but it is well below most state-of-the-art chemical networks. Another limitation of the code is the limited capability of the Kvaerno implicit stiff solver, which has reduced performance compared to well-established solvers \citep{Hindmarsh2005} and is not easily differentiable, unlike \diffrax{}. In this case, the technical difficulty was incorporating an optimization framework that treats the rate coefficient parameters as optimizable elements and vectorizes the optimization over a grid of models to simultaneously optimize the rate coefficients.

Another limitation that will impact computational resources (especially the time required per forward-modeling step) is the use of chemistry alongside hydrodynamics to address cooling and heating, and therefore computing the temperature evolution consistently \citep{Grassi2014}. In this paper, we consider chemistry as a post-processing stage. The critical limit here is the efficiency of the differentiable solver in \diffrax{}, which is not capable of the required speed, while another aspect is the lack of cooling and heating terms in \carbox{}. While the latter problem is shown to be easily solvable\footnote{see \url{https://github.com/GijsVermarien/carbox/pull/7}.}, the former requires a community effort to develop an effective solver with numerical efficiency comparable to \textsc{CVODE} \citep{Cohen1996} or \textsc{DLSODES} \citep{Hindmarsh2005}.

Finally, the present radiative transfer calculation only considers thin emission and therefore does not encounter the technical issues associated with optically thick radiative transfer. However, we have successfully tested a code version that includes these additional complications, but we will use it in future, more realistic applications. Analogously to hydrodynamics, the main issue concerns the unknown iteration steps, arising from the unknown convergence of the Lambda iterations used to solve the level population. This issue has been solved again with checkpointing, as discussed in \sect{sect:hydro} and \sect{sect:rt}. Apart from the optically thick issues, the code's main feature is its extensive vectorization, which allows us to compute emission lines with minimal numerical impact and full differentiability, though only in a 1D setup.

%--------------------------------------------------------------------
%--------------------------------------------------------------------
\section{Conclusions}\label{sect:conclusions}
We presented the first end-to-end differentiable pipeline that enables the solution of hydrodynamics, chemistry, and radiative transfer. This will allow us to use the observed spectra to optimize any desired model parameter at any stage of the computational pipeline. The hydrodynamic and the radiative transfer codes have been designed specifically for this work, while the chemical code is a modified version of \carbox{}. We showed that it is possible to leverage differentiability to optimize entire sets of chemical rate coefficients and to optimize the physical conditions of a shock to produce a target spectrum of \ce{HCO+}. Although still limited in complexity, this pipeline advocates using differentiable models as a promising research avenue for the future.
This work establishes a foundation for differentiable multiphysics modeling in astrophysics and provides a pathway toward next-generation simulation-based inference techniques.

%--------------------------------------------------------------------
\section{Tools employed}
The text of this paper has been refined with Grammarly, while parts of the code have been written using Copilot's inline completion in VS Code. The source code is not publicly available at this stage, as it is under active development. It will be released once a stable version is available.

%--------------------------------------------------------------------
%--------------------------------------------------------------------
\begin{acknowledgements}
TG, JEP, SS, and PC gratefully acknowledge the support of the Max Planck Society. SB acknowledges support from ANID Basal Project FB210003.
\end{acknowledgements}

\bibliographystyle{aa}
\bibliography{mybib}%,bibilio_extra}

%--------------------------------------------------------------------
%--------------------------------------------------------------------
\begin{appendix}

%--------------------------------------------------------------------
\section{Hydrodynamical equations}\label{appx:HLL}
The hydrodynamic equations are solved in conservative form,
\begin{equation}
 \frac{\partial \mathbf U}{\partial t} + \frac{\partial \mathbf F(\mathbf U)}{\partial z}=0\,,
\end{equation}
with
\begin{equation}
 \mathbf U=
 \begin{pmatrix}
  \rho\\
  \rho \varv\\
  E
 \end{pmatrix},
 \qquad
 \mathbf F=
 \begin{pmatrix}
  \rho \varv\\
  \rho \varv^2+p\\
  v(E+p)
  \end{pmatrix}\,,
\end{equation}
and
\begin{equation}
  E=\frac{p}{\gamma-1}+\frac12\rho \varv^2\,,
\end{equation}
where $z$ is the spatial coordinate, $\rho$ is the mass density, $\varv$ is the $z$ (and only) component of the velocity, $p$ is the pressure, $E$ is the energy density, and $\gamma$ is the adiabatic index. 

Using a finite-volume discretization on a uniform cell grid, the semi-discrete update is
\begin{equation}
\frac{\dd\mathbf U_i}{\dd t} = -\frac{\mathbf F_{i+1/2}-\mathbf F_{i-1/2}}{\Delta z}\,.
\end{equation}

The intercell fluxes are computed with the HLL approximate Riemann solver,
\begin{equation}
 \mathbf F_{\rm HLL} = \frac{S_R\mathbf F_L - S_L\mathbf F_R + S_LS_R(\mathbf U_R-\mathbf U_L)}{S_R-S_L}\,,
\end{equation}
with
\begin{equation}
 S_{\rm L}=\min(v_{\rm L} - c_{\rm s, L}, v_{\rm R} - c_{\rm s, R}),
 \qquad
 S_{\rm R}=\max(\varv_{\rm L} + c_{\rm s, L},\varv_{\rm R} + c_{\rm s, R})\,,
\end{equation}
where the left and right components are considered, and where $c_{\rm s}^2\,\rho=\gamma\,p$ is employed to compute the speed of sound $c$.

In addition to these, we use $\rho = n\,\mu$, where $n$ is the gas number density and $\mu$ its mean molecular weight, and $p=n\,k_{\rm B} T$, where $k_{\rm B}$ is the Boltzmann constant, while $T$ is the gas temperature.

The code allows three types of boundary conditions: Dirichlet (constant derivatives), fixed (zero derivatives), or periodic. We use Dirichlet boundary conditions unless otherwise specified.

% -----------------------------
\section{Rate coefficient fitting}\label{appx:testcarbox}
We use the library \optax{} \citep{Optax2020} to optimize the rate coefficients for a given set of time-dependent abundances $n(t; T, \zeta)$ at different temperatures and cosmic ray ionization rates. Our target quantity is a $N_t\times N_T \times N_\zeta$ matrix, where $N_t=50$ is the number of timesteps ($1-10^5$~yr, log-spaced), $N_T=5$ the number of temperatures (10-300\,K, log-spaced), and $N_\zeta$ the number of cosmic rays values ($10^{-18}-10^{-14}$\,s$^{-1}$, log-spaced). 

The optimization target $N_t\times N_T \times N_\zeta$ matrix is constructed by running \carbox{} on a grid of $T$ and $\zeta$ values for a specific set of $\alpha$, $\beta$, and $\gamma$ (i.e., the values found in the KIDA database, \citealt{Wakelam2024}). For numerical stability, this matrix is normalized in the logarithmic space, and the loss is constructed by minimizing the $L_2$-norm of the difference between the target and the computed matrix. The initial parameters are also normalized to their expected ranges in log space, except for $\beta$ and the photochemistry $\gamma$, which are normalized in linear space. We found that the exact normalization range is not relevant to the optimization, whereas the choice of logarithmic or linear space is crucial. We then randomize the parameters to have a completely different set of initial conditions. Thanks to automatic differentiation, \optax{} (using an Adam optimizer with a learning rate of $0.03$, \citealt{Kingma2014}) converges after approximately 1000 iterations, which takes approximately 20 minutes on an NVIDIA Quadro RTX 4000 (8GB, 2018 version), depending on the initial conditions. The number of iterations is significantly reduced if only a subset of reactions is optimized, with the 63-parameter optimization representing an extreme case.

\begin{figure}
\centering
    \includegraphics[width=0.48\textwidth]{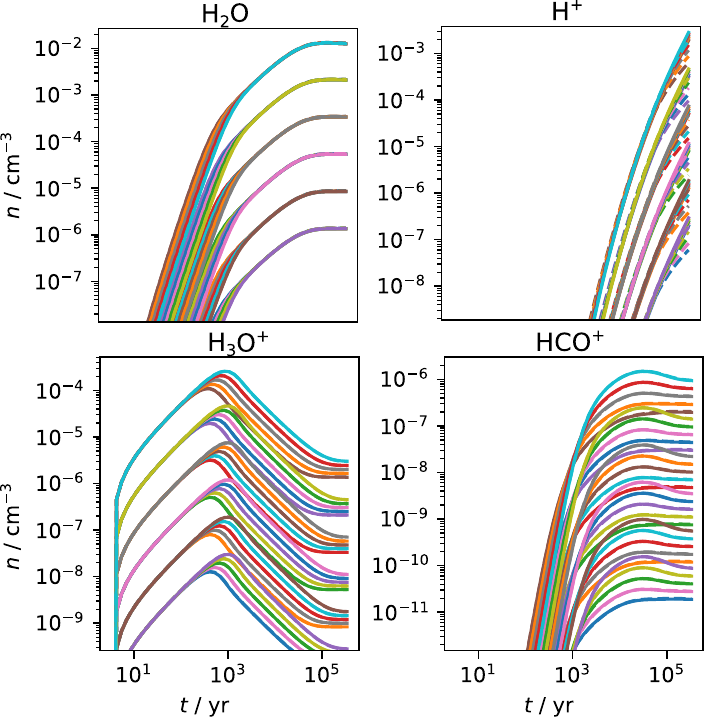}
    \caption{Time evolution of the abundances of selected chemical species (4 out of 16). Each panel represents a chemical species, while each trajectory is for a specific $T$ and $\zeta$ combination. The solid lines are the target solutions, while the dashed line is the optimized solution, which, in most cases, perfectly overlaps, apart from \ce{H+}, which shows the largest difference among all 16 molecules.}
        \label{fig:carbox}
\end{figure}

We report the results of this optimization in \fig{fig:carbox}, where each panel represents the time-dependent abundance of a chemical species, and each trajectory is computed at a given $T$ and $\zeta$ combination. The dashed lines are the optimized model, while the solid lines are the ground truth. 

% ------------------------------
\section{Radiative transfer benchmarks}\label{appx:testrt}
\begin{figure}
\centering
    \includegraphics[width=0.48\textwidth]{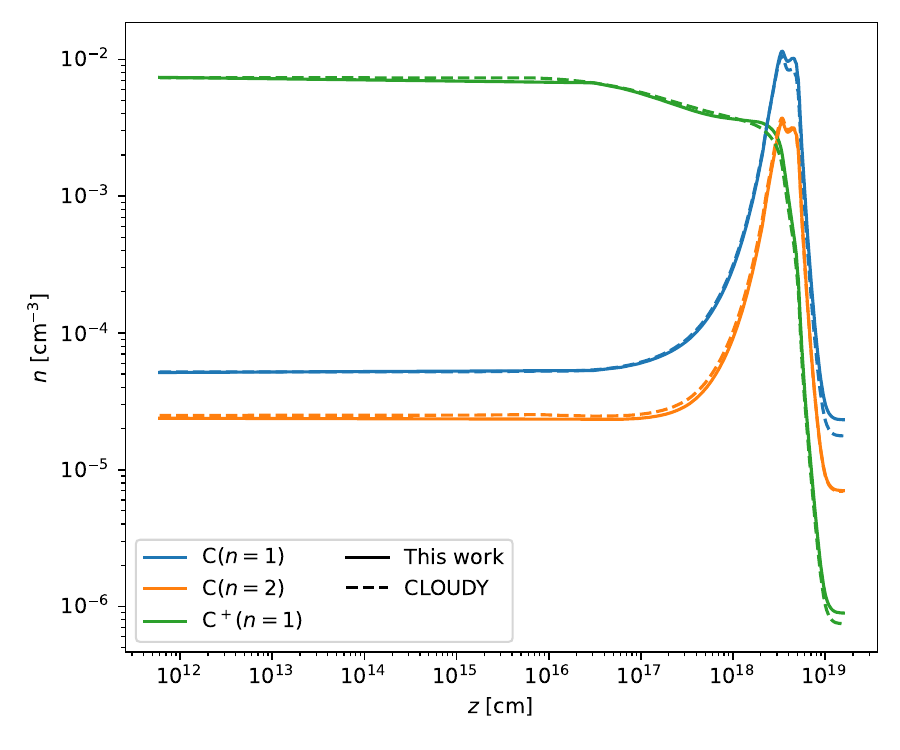}
    \caption{PDR population level of C and \ce{C+}, where the ground levels ($n=0$) are omitted for clarity. The dashed lines are for \textsc{Cloudy}, while our model has solid lines.}
        \label{fig:test_pdr}
\end{figure}

To test our radiative transfer module, we first verified the level-population calculations against the results of the PDR benchmark by \citet{Rollig2007}. In particular, we used the output of \textsc{Cloudy} for the F1 model (constant temperature and density, i.e., $T=50$\,K, $n=10^3$\,cm$^{-3}$). Giving that the emission of a specific transition $u\to l$ is $I_{ul}=n_u\,A_{ul}\,(E_u-E_l)$, we inferred the population level $n_i$ from the emission of C at 370 and 610\,$\mu$m, and \ce{C+} at 158\,$\mu$m as a function of spatial position within the slab. We employed the chemical composition computed by \textsc{Cloudy} (i.e., $n_{\rm C}$ and $n_{\ce{C+}}$) and compared the inferred level population of \textsc{Cloudy} with our code. For C, the colliders are H, \ce{H+}, \ce{e-}, and \ce{H2}, while for \ce{C+} we have H, \ce{e-}, and \ce{H2}, where in both cases we assume an ortho-to-para ratio of 1, but the exact ratio has no relevant impact on the final result.  The results are in \fig{fig:test_pdr}, where the level populations overlap.

\begin{figure}
\centering
    \includegraphics[width=0.48\textwidth]{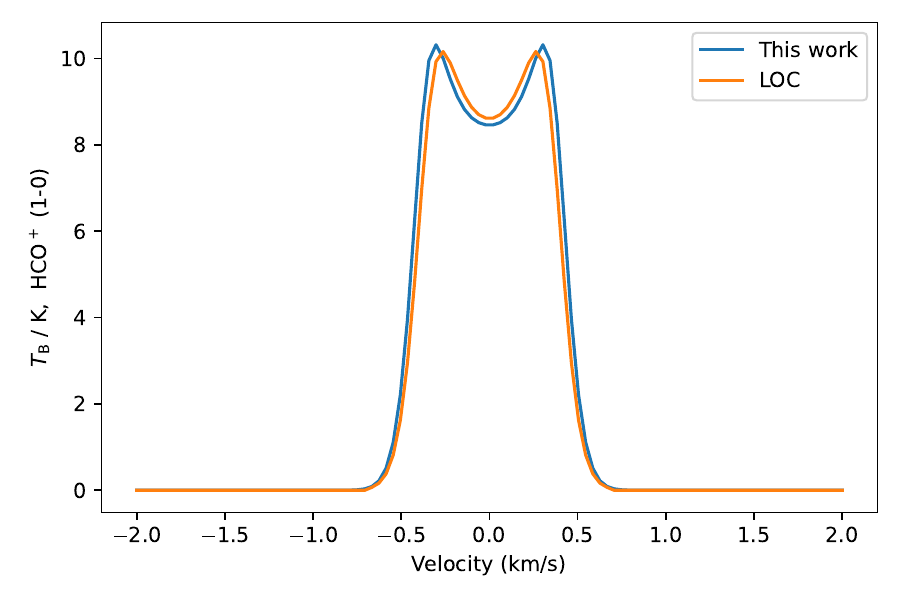}
    \caption{Emission of the $1\to0$ transition of \ce{HCO+} from our code (blue line), and from LOC (orange line).}
        \label{fig:test_loc}
\end{figure}

To test the emission and absorption components, we instead compared our emission spectrum with that produced by LOC. The latter is a code that requires defining a 3D structure. We therefore constructed a $64\times128\times64$ cartesian box, where the $y$ component corresponds to the line of sight (LOS). The dimension of the box along the LOS is $10^6\,$au, and half of the size in the other two dimensions, with a total density of \ce{H2} of $10^4$\,cm$^{-3}$, and \ce{HCO+} fraction of $10^{-9}$, both constant in the box. The temperature is 30\,K, and the microturbulence is $0.2$\,km\,s$^{-1}$. We assume no background radiation in this test. We noted that, due to the geometric differences between the two codes, the level populations differ, especially at the boundary conditions. To avoid geometric effects and test the emission/absorption machinery only, we therefore provided the level populations calculated by LOC to our code. The brightness temperature of the $1\to0$ transition of \ce{HCO+} is reported in \fig{fig:test_loc}.
% -----------------------------
\section{Loss and convergence}\label{appx:loss}
\begin{figure}
\centering
    \includegraphics[width=0.48\textwidth]{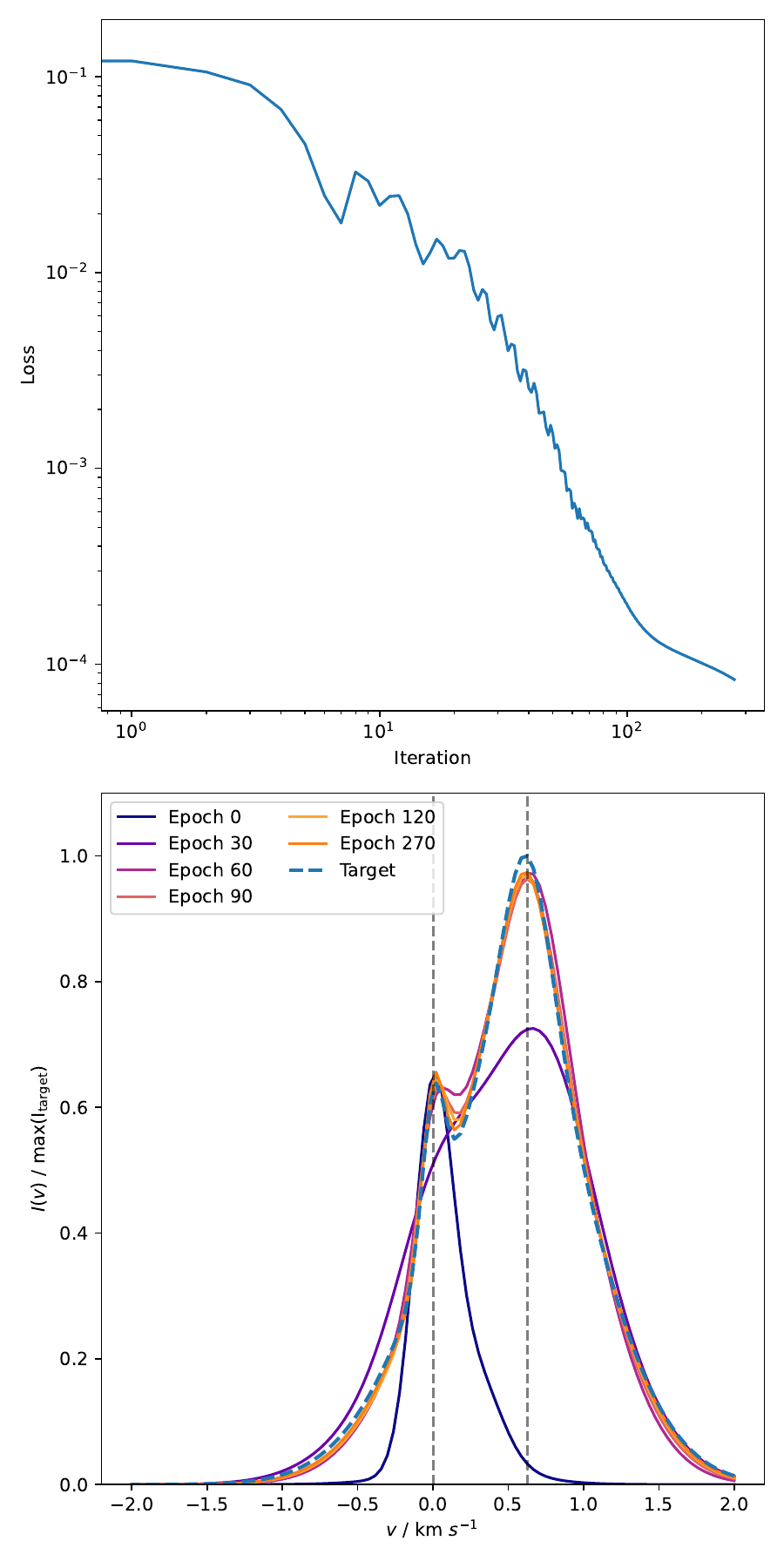}
    \caption{Top panel: the loss function as a function of iterations. Bottom panel: the observed intensity over epochs. During the early epochs, the optimization shifts the distribution to higher velocities. At later epochs, the shape morphs into the asymmetric target distribution. The dashed curve is the target spectrum, while the vertical dashed lines indicate the peaks discussed in the text. }
        \label{fig:loss}
\end{figure}

In \fig{fig:loss}, we report the loss evolution (top) and the resulting spectrum every 30 epochs up to 120, followed by the final result at 270 epochs (bottom). We note that the learning rate leads to large oscillations during gradient descent, but convergence is relatively straightforward. We also noted that not all randomized initial conditions yield a smooth gradient descent, and in some cases, the optimizer's learning rate needs to be fine-tuned. The bottom panel shows that convergence toward the target spectrum is relatively fast, producing viable results by around 60 epochs. We also note that the final result is not perfectly reproduced, as evidenced by a small difference in the main peak that is not recovered after 270 epochs.

% ------------------------------
\section{Chemical network}\label{appx:network}
In \tab{tab:rates}, we report the reactions and the corresponding rate coefficients.

\begin{table}
    \small
    \caption{Reaction rate coefficients used in this work.}
    \centering
    \begin{tabular}{lllll}
        \hline
        &Reaction & & & Rate coefficient ($k$)\\
        \hline

1 & \ce{O} + \ce{CR} &$\to$& \ce{O+} + \ce{e-} & $2.80\,\zeta$\\
2 & \ce{C} + \ce{CRP} &$\to$& \ce{C+} + \ce{e-} & $2.62\,\zeta$\\
3 & \ce{CO} + \ce{CR} &$\to$& \ce{C} + \ce{O} & $5.00\,\zeta$\\
4 & \ce{H2} + \ce{CR} &$\to$& \ce{H} + \ce{H} & $0.01\,\zeta$\\
5 & \ce{O+} + \ce{H2} &$\to$& \ce{OH+} + \ce{H} & $1.60\times10^{-9}$\\
6 & \ce{OH+} + \ce{H2} &$\to$& \ce{H2O+} + \ce{H} & $1.00\times10^{-9}$\\
7 & \ce{H2O+} + \ce{H2} &$\to$& \ce{H3O+} + \ce{H} & $6.10\times10^{-10}$\\
8 & \ce{H3O+} + \ce{e-} &$\to$& \ce{H2O} + \ce{H} & $1.10\times10^{-7}T_3^{-0.50}$\\
9 & \ce{H2O+} + \ce{e-} &$\to$& \ce{OH} + \ce{H} & $8.60\times10^{-8}T_3^{-0.50}$\\
10 & \ce{H2O+} + \ce{e-} &$\to$& \ce{O} + \ce{H2} & $3.90\times10^{-8}T_3^{-0.50}$\\
11 & \ce{OH+} + \ce{e-} &$\to$& \ce{O} + \ce{H} & $6.30\times10^{-9}T_3^{-0.48}$\\
12 & \ce{O+} + \ce{e-} &$\to$& \ce{O} + $\gamma$ & $3.40\times10^{-12}T_3^{-0.63}$\\
13 & \ce{C+} + \ce{e-} &$\to$& \ce{C} + $\gamma$ & $4.40\times10^{-12}T_3^{-0.61}$\\
14 & \ce{C} + \ce{OH} &$\to$& \ce{CO} + \ce{H} & $1.15\times10^{-10}T_3^{-0.34}$\\
15 & \ce{C+} + \ce{OH} &$\to$& \ce{CO+} + \ce{H} & $5.67\times10^{-10}$\\
16 & \ce{C+} + \ce{OH} &$\to$& \ce{CO+} + \ce{H} & $2.40\times10^{-9}T_3^{-0.50}$\\
17 & \ce{CO+} + \ce{H} &$\to$& \ce{CO} + \ce{H+} & $4.00\times10^{-10}$\\
18 & \ce{HCO+} + \ce{e-} &$\to$& \ce{CO} + \ce{H} & $2.80\times10^{-7}T_3^{-0.69}$\\
19 & \ce{CO+} + \ce{H2} &$\to$& \ce{HCO+} + \ce{H} & $7.28\times10^{-10}$\\
20 & \ce{H+} + \ce{e-} &$\to$& \ce{H} + $\gamma$ & $3.50\times10^{-12}T_3^{-0.70}$\\
21 & \ce{H} + \ce{H} &$\to$& \ce{H2} & $2.12\times10^{-17}$ (dust)\\
22 & \ce{H2} + $\gamma$ &$\to$& \ce{H} + \ce{H} & $5.68\times10^{-11}\exp(-4.18\,A_{\rm V})$\\
23 & \ce{C} + $\gamma$ &$\to$& \ce{C+} + \ce{e-} & $3.39\times10^{-10}\exp(-3.76\,A_{\rm V})$\\
24 & \ce{CO} + $\gamma$ &$\to$& \ce{C} + \ce{O} & $2.43\times10^{-10}\exp(-3.88\,A_{\rm V})$\\
25 & \ce{H2O} + $\gamma$ &$\to$& \ce{OH} + \ce{H} & $7.72\times10^{-10}\exp(-2.63\,A_{\rm V})$\\
        \hline
    \end{tabular}
\tablefoot{$T_3=T/300$\,K. CR are cosmic rays, CRP photons from cosmic rays-induced excitations, and $\gamma$ photons. Reaction rates are from \cite{Anicich1986,Amano1990,Jensen2000,Woon2009,Zanchet2009,Harada2010,Wakelam2024}.}\label{tab:rates}
\end{table}

\end{appendix}

\label{LastPage}
\end{document}